\documentclass[prl,showpacs,twocolumn,aps]{revtex4}
\usepackage{rotating}
\usepackage{epsfig}
\newcommand{\be}{\begin{eqnarray}}
\newcommand{\bff}{\rm}
\newcommand{\ee}{\end{eqnarray}}
\newcommand{\bea}{\begin{eqnarray}}
\newcommand{\eea}{\end{eqnarray}}

\begin{document}
\title{Self-Consistent Modification To The Electron
Density Of States Due To Electron-Phonon Coupling In Metals}
\author{F. Do\u gan and F. Marsiglio}

\affiliation{Department of Physics, University of Alberta, Edmonton,
Alberta, Canada T6G 2J1}

\date{\today}
\begin{abstract}
The "standard" theory of a normal metal consists of an effective
electron band which interacts with phonons and impurities. The
effects due to the electron-phonon interaction are often delineated
within the Migdal approximation; the properties of many simple metals
are reasonably well described with such a description. On the other hand,
if the electron-phonon interaction is sufficiently strong, a polaron
approach is more appropriate. The purpose of this paper is to examine
to what degree the Migdal approximation is self-consistent, as the
coupling strength increases. We find that changes in the electron density
of states become significant for very large values of the coupling
strength; however, there is no critical value, nor even a crossover regime
{\bff where the Migdal approximation has become inconsistent}.
Moreover, the extent to which the electron band
collapses is strongly dependent on the detailed characteristics of the phonon
spectrum.

\end{abstract}
\maketitle
\date{\today}


\section{INTRODUCTION}
\label{sec:int}

The Migdal approximation for the electron self energy due to the
electron-phonon interaction consists of neglecting vertex corrections.
This procedure was first justified by Migdal \cite{migdal58} based
on an approximate treatment of the first-order vertex correction.
He found that the correction to the bare vertex is of order
$O(\lambda{\omega_D \over \epsilon_F})$, where $\lambda$ is the
dimensionless electron-phonon coupling constant, $\omega_D$ is
the typical phonon frequency, and $\epsilon_F$ is the electron
Fermi energy. If we ignore the factor of $\lambda$ \cite{note1},
then the ratio $\omega_D / \epsilon_F$ is generally very small in a metal.

Subsequently, Engelsberg and Schrieffer \cite{engelsberg63}
performed numerical calculations of the self energy and spectral
function, based on the Migdal approximation \cite{note2}. The
result is found in several reviews and texts
\cite{allen82,nakajima80}, and we quote here the main results. The
electron self energy is given by a frequency dependent, momentum
independent function, \be \Sigma(\omega+i\delta) = \int_0^\infty
d\nu \, \, \alpha^2F(\nu) \biggl[ \, \,
-2\pi i (n(\nu) + 1/2 ) +\nonumber\\
\psi \bigl( {1 \over 2} + i{\nu - \omega \over 2\pi T} \bigr) -
\psi \bigl( {1 \over 2} - i{\nu + \omega \over 2\pi T} \bigr) \,
\, \biggr]\label{self1} \ee where $\psi (x)$ is the digamma
function \cite{abramowitz64,allen82} and the entire expression has
been written for a frequency just above the real axis,
$\omega+i\delta$. In Eq. (\ref{self1}) $n(\nu)$ is the bose
function, and $\alpha^2F(\nu)$ is the electron-phonon spectral
function. A truly self-consistent approach would require, amongst
other things, a self-consistent correction to the phonon spectrum,
due to the interaction with electrons. Migdal estimated this
correction, and found that the phonon frequencies are
renormalized, {\bff and an instability is encountered as the bare
coupling strength increases. However, we are adopting a more
phenomenological approach here. The common practice
\cite{scalapino69} is to take information concerning the phonons
from experiment as input into the theory for the electrons. The
justification for this comes from experiment, where well-defined
phonons are observed in neutron scattering experiments
\cite{brockhouse62}, for example. Since these are used in the
theory for the electron properties, it would be incorrect to
compute renormalizations for the phonons.} We follow this
philosophy in everything that follows \cite{note3}.

Nonetheless, Eq. (\ref{self1}) was obtained with a number of other simplifying
assumptions and approximations. In particular the self energy of the electron
is determined by an infinite set of diagrams in which phonon lines do not
cross; these can be summarized by the diagram in Fig. 1, where the full
electron Green function, $G({\bf k},\omega+i\delta) \equiv 1/(\omega + i\delta -
\epsilon_{\bf k} - \Sigma({\bf k},\omega+i\delta))$, is required, and of course
depends on the very self energy that we are trying to calculate. Yet
Eq. (\ref{self1}) shows no sign of self-consistency. The reason was noted
already in Ref. \cite{migdal58} and arises because the bandwidth is assumed
to be essentially infinite compared to the typical phonon energy. Then, the
nested diagrams which arise from iterating the equation in Fig. 1 all
contribute zero, and the same result is obtained by simply replacing the
full electron Green function in the figure with the non-interacting Green
function.
\begin{figure}[tp]
\begin{center}
\epsfig{figure=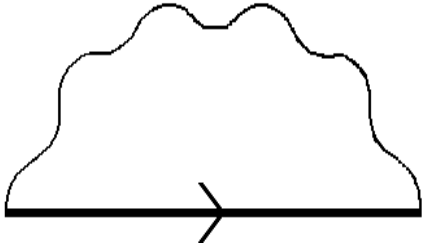,height=1.5in,width=3.0in} \caption{ Diagram
for the electron self energy. Note that the self-consistent
electron Green function (heavy solid line) is used in this
calculation. }
\end{center}
\end{figure}

Engelsberg and Schrieffer \cite{engelsberg63} relaxed the assumption
of infinite bandwidth \cite{mitrovic83}, but used
realistic values for the phonon frequency and Fermi energy for materials known
at that time. They found only very small effects. More recently, Alexandrov
et al. \cite{alexandrov87} readdressed the question of the impact of a finite
(i.e. not infinite) bandwidth on the electron properties, and adopted much
more extreme values of the ratio of the typical phonon frequency to Fermi energy,
$\omega_D/\epsilon_F$ (referred to hereafter as the frequency ratio).
They concluded that the Migdal approximation breaks down for coupling strengths
that exceed unity.

In this paper we wish to assess this conclusion, by examining the effect of a
more realistic phonon spectral function. Alexandrov et al. \cite{alexandrov87}
used an Einstein spectrum to simplify the calculation. This spectrum is,
of course, singular, and it is perhaps not too surprising if singular behaviour
in the electron properties results. We will first outline the problem as posed
by Alexandrov et al. and demonstrate that singular behaviour exists for any
coupling strength. We argue that this behaviour does not necessarily invalidate
the calculation, as tacitly assumed in Ref. \cite{alexandrov87}. Instead those
authors properly focused on the more global behaviour of the electron density of
states (EDOS), as a function of increased coupling strength. We examine in much
more detail this global behaviour as a function of the parameters in the problem,
so that a more quantitative assessment of the breakdown can be obtained. In
particular we examine the dependency of the band collapse on the frequency ratio, the
electron-phonon coupling strength, the presence of a secondary band, the shape of
the bare band (Lorentzian versus square), and finally, the shape of the electron
phonon spectral function, $\alpha^2F(\nu)$. For convenience $\alpha^2F(\nu)$ is
modified from an Einstein spectrum to a Lorentzian. In this way a single parameter
(the width of the spectrum) controls its shape. We find that there is no clear
transition or even crossover to a regime where the Migdal approximation has become
inconsistent. {\bff Nonetheless, this conclusion is not meant to imply that this
calculation shows that the Migdal approximation is accurate in the intermediate
or strong coupling regime. As will be summarized in the final section, other work
suggests that this is not the case. Our calculation merely shows that within the
Migdal framework, a signal of this potential breakdown does {\bf not} occur, if
a broad phonon spectrum is used.}

\section{THE SELF-CONSISTENT MIGDAL APPROXIMATION}
\label{sec:mig}

The self-consistent Migdal approximation results in the following
equation for the electron self energy $\Sigma(\omega+i\delta)$:
\be \Sigma(\omega+i\delta) = \int_0^{\infty}d\nu \, \alpha^2F(\nu)
\, \hspace{1.4in}\nonumber\\\int_{-\infty}^{\infty}d\omega^\prime
\, \frac{N(\omega^\prime)}{N_\circ(0)} \biggl[\frac{n(\nu) +
f(-\omega^\prime)}{\omega+i\delta - \nu - \omega^\prime} +
\frac{n(\nu) + f(\omega^\prime)}{\omega+i\delta + \nu -
\omega^\prime}\biggr] \label{self2} \ee where
\begin{equation}
N(\omega)=\int_{-\infty}^{\infty}d\epsilon \,
N_\circ(\epsilon)A(\epsilon,\omega).
\label{self3}
\end{equation}
In these equations $f(\omega)$ and $n(\nu)$ are the Fermi and Bose distribution
functions, respectively, $N_\circ(\epsilon)$ is the non-interacting (bare)
electron density of states, and $N(\omega)$ is the self-consistently calculated
electron density of states. The electron spectral function, $A(\epsilon,\omega)$
 is given by
\be
A(\epsilon,\omega) \equiv -{1 \over \pi} {\rm Im} G(\epsilon,\omega+i\delta),
\label{spect}
\ee
where the single electron Green function is given by
\be
G(\epsilon,\omega + i\delta) = {1 \over \omega+ i\delta - \epsilon
- \Sigma(\omega + i \delta)}.
\label{green}
\ee
Notice that we have tacitly assumed that the electron self energy is independent
of momentum (i.e. independent of $\epsilon$). The arguments that justify this
simplification are
provided, for example, in Ref. \cite{allen82}.

In Eq. (\ref{self3}), if $N_\circ(\epsilon)$ is taken to be a
constant ($=N_\circ(0)$), extending over all energies, then using
the fact that $\Sigma$ is independent of momentum (i.e.
$\epsilon$), we obtain $N(\omega) = N_\circ(0)$. Thus, the
standard approximation, that the Fermi energy and bandwidth are
large energies compared to the phonon energy (so that we can
neglect the former and simply integrate from $-\infty$ to
$\infty$) leads to an electron density of states which is
unmodified by the electron-phonon interaction. This is true even
though the self energy has a non-trivial frequency dependence
\cite{note4}. When a more realistic bare electron density of
states is used, then Eq. (\ref{self3}) leads to an altered EDOS.
For example, with $N_\circ(\epsilon) = N_\circ(0) \theta(D/2 -
|\epsilon|)$, i.e. a constant over a limited energy range, Eq.
(\ref{self3}) gives \be {N(\omega) \over N_\circ(0)} =
\frac{1}{\pi}\biggl[ \arctan\bigl(\frac{D/2 + Re\Sigma(\omega + i
\delta) - \omega}
{|Im\Sigma(\omega + i \delta)|}\bigr) +\nonumber\\
\arctan\bigl(\frac{D/2 - Re\Sigma(\omega + i \delta) + \omega}
{|Im\Sigma(\omega + i \delta)|}\bigr) \biggr].\label{den_con} \ee
If a Lorentzian form is used the bare density of states is given
by \be N_\circ(\epsilon) = N_\circ(0) {(D/2)^2 \over \epsilon^2 +
(D/2)^2}. \label{denb_lor} \ee In either case $N_\circ(0)$ is the
density of states at the Fermi level and $D$ is the full
bandwidth, defined in an obvious way in the case of the constant
case, and as the full width at half maximum in the Lorentzian
case. For this latter case, Eq. (\ref{self3}) gives
\cite{alexandrov87} \be {N(\omega) \over N_\circ(0)} =
\frac{D/2(D/2+|{\rm Im}\Sigma(\omega + i \delta)|)} {({\rm
Re}\Sigma(\omega + i \delta)-\omega)^2 + (|{\rm Im}\Sigma(\omega +
i \delta)|+D/2)^2}. \label{den_lor} \ee Eq. (\ref{den_lor}) or
(\ref{den_con}) is required to self-consistently calculate the
electron self energy given by Eq. (\ref{self2}). In what follows
we further simplify the calculation by adopting $T=0$, as in Ref.
\cite{alexandrov87}. This simplifies Eq. (\ref{self2}) since
$f(\omega) \rightarrow \theta(-\omega)$ and $n(\nu) \rightarrow 0$
in this limit. Also, we will adopt particle-hole symmetry
throughout this paper. The final equations, separated out into
their real and imaginary parts, are given by \be
{\rm Re} \Sigma(\omega + i\delta) = \hspace{2.5in}\nonumber\\
\int_0^\infty \, d\nu \, \alpha^2F(\nu) \biggl[
\int_{0}^{\infty}d\omega^\prime \, {N(\omega^\prime) \over
N_\circ(0)} {2\omega \over \omega^2 - (\nu + \omega^\prime)^2}
\biggr] \label{resig} \ee and \be {\rm Im} \Sigma(\omega+i\delta)
= -\pi \int_0^\infty \, d\nu \, \alpha^2F(\nu) \biggl[
{N(\omega-\nu) \over N_\circ(0)} \theta(\omega-\nu) +
\nonumber\\{N(\omega+\nu) \over N_\circ(0)} (1-\theta(\omega +
\nu)) \biggr].\hspace{.5in} \label{imsig} \ee As mentioned in the
introduction, the primary purpose of this paper is to explore the
consequences of a non-singular phonon spectrum. For simplicity we
will adopt a Lorentzian spectrum for the phonons, given by \be
\alpha^2F(\nu) = \hspace{2.5in}\nonumber\\\frac{\lambda^\prime
\omega_E}{2\pi}\biggl(
\frac{\delta}{(\nu-\omega_E)^2+\delta^2}-\frac{\delta}{\eta^2+\delta^2}
\biggr) \theta(\eta-|\omega_E-\nu|),  \label{phon_lor} \ee where
the subtracted term ensures that the spectrum is continuous
everywhere, particularly at the endpoints, and $\eta$ is the full
width of the spectral function (less than $\omega_E$). As the
parameter $\delta$ approaches zero, this spectrum approaches an
Einstein spectrum centered at $\omega = \omega_E$, and
$\lambda^\prime \rightarrow \lambda$. Fig.~2 shows several spectra
for different values of $\delta$.
\begin{figure}[tp]
\begin{center}
\begin{turn}{-90}
\epsfig{figure=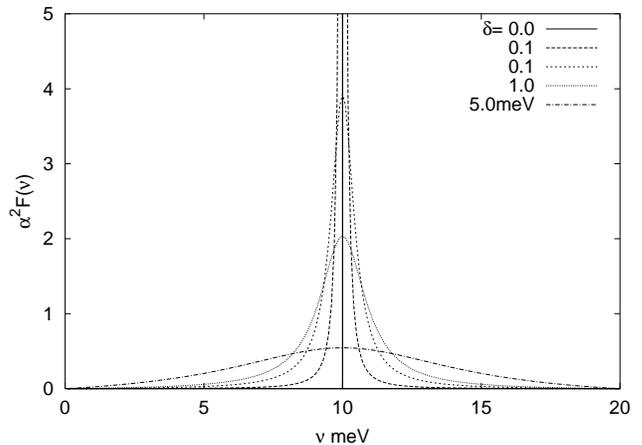,height=3.4in}
\end{turn}
\caption{The sequence of phonon spectral functions used to model
broadening. The curves are truncated Lorentzians as described in
the text with widths as labelled. }
\end{center}
\end{figure}

As a technical aside, simplifications occur for the Einstein
spectrum, where $\alpha^2F(\nu) = {\lambda \omega_E \over 2}
\delta(\nu - \omega_E)$. As is apparent from Eq. (\ref{imsig}),
${\rm Im}\Sigma(\omega + i \delta)$ becomes simply related to the
self-consistent EDOS. The real part of the self energy becomes
singular, as is evident when Eq. (\ref{resig}) is rewritten for an
Einstein phonon spectrum as \be {\rm Re}\Sigma(\omega + i \delta)
= \hspace{2.5in}\nonumber\\{\lambda \omega_E \over 2} \biggl[
\int_{0}^{\infty}d\omega^\prime \, \bigl({(N(\omega^\prime) -
N(\omega-\omega_E) \over N_\circ(0)}\bigr) {2\omega \over \omega^2
- (\omega_E + \omega^\prime)^2} \nonumber\\+ {N(\omega-\omega_E)
\over N_\circ(0)} \, {\rm ln}|{\omega - \omega_E \over \omega +
\omega_E}| \biggr], \hspace{.5in}\label{resig2} \ee and the
logarithmic singularity is now explicit. Using an EDOS that is
constant with infinite bandwidth gives zero for the
$\omega^\prime$ integral in Eq. (\ref{resig2}), and we recover the
`standard' result \cite{engelsberg63} for the electron self
energy.

\section{RESULTS}
\label{sec:res}

\noindent {\it Einstein phonon spectrum}

It is clear from the discussion in the previous section that there is
a simple scaling relation amongst the energies in the problem. Nonetheless
we will use real units, and the reader can scale the results to other
energy scales, if so desired. We begin with $\omega_E = 10$ meV, and
use a bandwidth 10X this amount, i.e. $D = 10\omega_E = 100$ meV. For
definiteness we use $\lambda = 2$, which is considered very strong coupling
(Pb, for example, has $\lambda \approx 1.5$), and $N_\circ(0) = 1/D$.
In Fig. 3 we plot the (a) real and (b) imaginary parts of the electron
self energy as a function of frequency for these parameters. We adopt a
Lorentzian shape for the bare EDOS. Three curves are shown; one is for the
standard theory, where an infinitely wide band with constant density of states
is assumed, the second is for the non-self-consistent result, where the
bare EDOS ($N_\circ(\omega)$) is substituted for $N(\omega)$ on
the right hand side of Eqs. (\ref{resig2},\ref{imsig}), and the third
represents the full self-consistent solution to these same equations,
using Eq. (\ref{den_lor}) in addition.
\begin{figure}[tp]
\begin{center}
\begin{turn}{-90}
\epsfig{figure=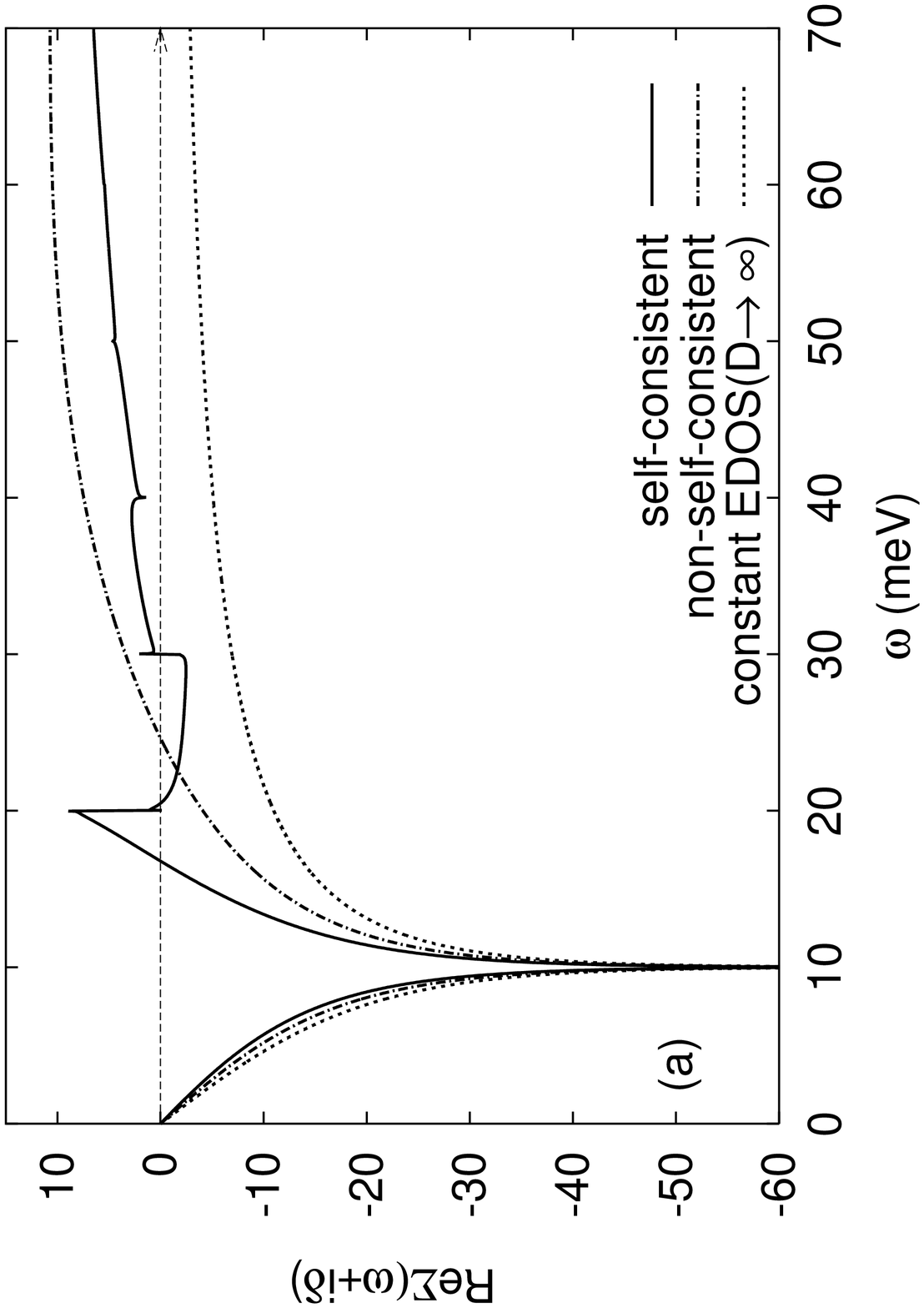,height=2.4in}
\epsfig{figure=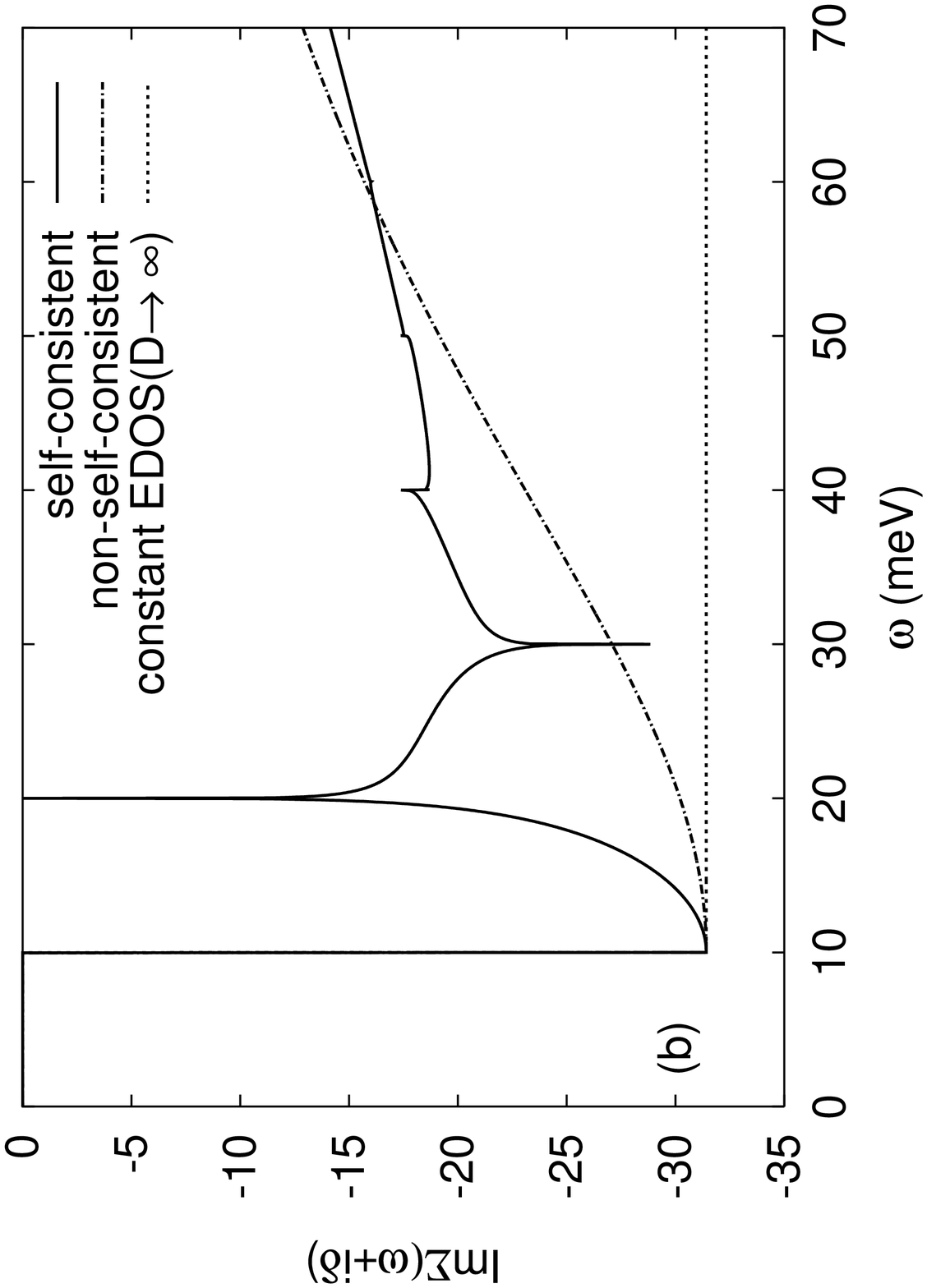,height=2.4in}
\epsfig{figure=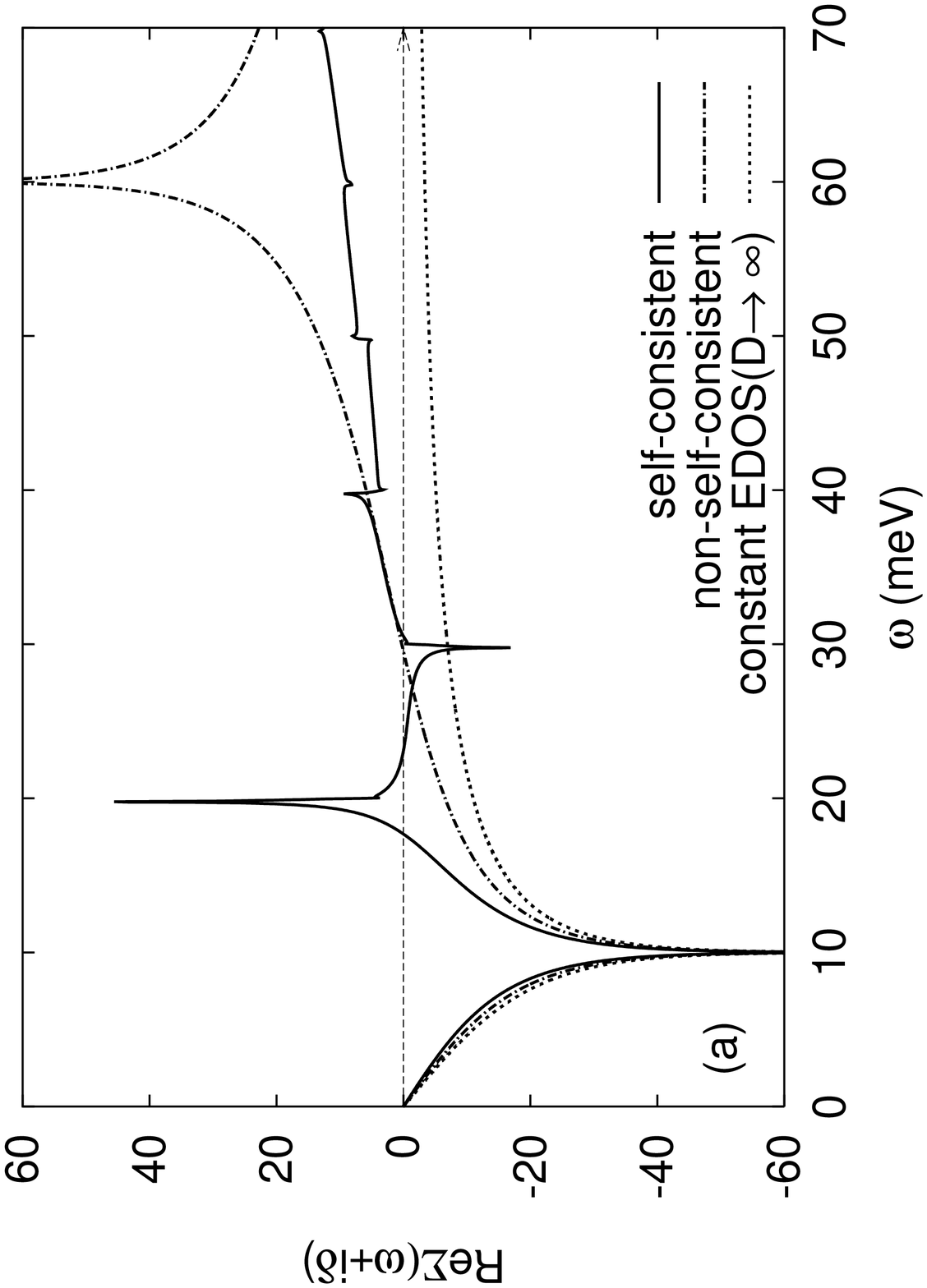,height=2.4in}
\epsfig{figure=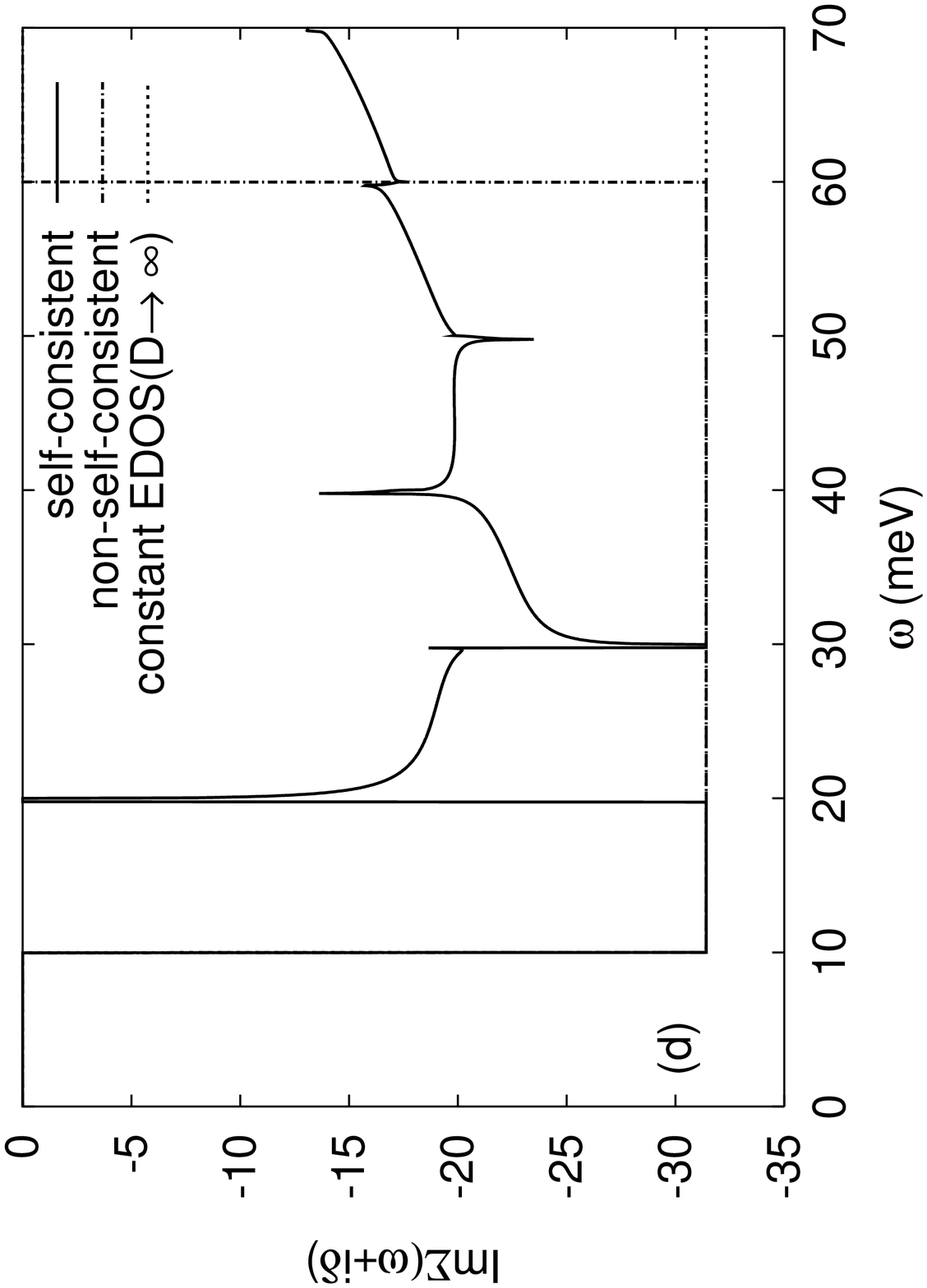,height=2.4in}
\end{turn}
\caption{The (a) real and (b) imaginary parts of the electron self
energy as a function of frequency, using an Einstein spectrum for
the phonons and a Lorentzian for the bare EDOS, with $\omega_E/D =
1/10$. The curves are computed with $\omega_E = 10$ meV, and
$\lambda = 2$. The non-self-consistent result uses the diagram
shown in Fig. 1 with the fully interacting electron Green function
replaced with the non-interacting electron Green function. Also
shown is the `standard' result with infinite electron bandwidth
(short-dashed curve). Note the singularity in the real part of the
self energy, which occurs at all levels of approximation. Parts
(c) and (d) show similar results obtained with a constant bare
EDOS with bandwidth $D$. An additional singularity occurs at
$\omega = D + \omega_E$ and is due to the abrupt band-edge in this
model. Note that the imaginary parts of the self energy, (b) and
(d), are zero up to $\omega_E$, and then turn on abruptly, because
the phonon is a $\delta$-function. For the constant density of
states (d) this imaginary part of the self energy extends to
infinity in the case where the bandwidth is infinite (dotted
curve). For a finite bandwidth (dashed curve), the imaginary part
of the self energy extends to $D+\omega_E$. }
\end{center}
\end{figure}

In parts (c) and (d) we plot the same quantities, calculated this time
for a bare EDOS which is constant between $-D/2$ and $D/2$. What is clear
from these plots is that a singularity in the real part of the self energy
exists at the Einstein frequency, regardless of the particular
approximation used. In fact, as Eq. (\ref{resig2}) makes clear, the
singularity is logarithmic, and exists {\it regardless of the value of
$\lambda$}. Thus, even for $\lambda \approx 0.01$, as long as it is
nonzero, the Migdal approximation results in a logarithmic singularity
in the self energy. That this is not a serious problem is hinted at by
Eq. (\ref{resig}), where one can see that, as long as a broader phonon spectrum
is used, the logarithmic singularity will be integrated to a non-singular
result. Furthermore, as mentioned earlier, even in the case of an Einstein
phonon spectrum, the electron density of states will remain unaltered when
the `standard' approximation of infinite bandwidth is used. The point of
Ref. \cite{alexandrov87} was, however, that this is not the case when
a bare EDOS with a non-infinite bandwidth is used. The EDOS is plotted in
Fig. 4, again using the same parameters as in Fig. 3, and for the same
levels of approximation (the bare EDOS
is also included for reference). In Fig. 4a (b) we use a bare
EDOS which is Lorentzian (constant) with bandwidth $D$. In both
cases the singularity manifests itself in the final EDOS in both
the non-self-consistent and self-consistent Migdal approximations. Thus,
it would appear that the EDOS has collapsed, and the effective bandwidth
is of order $2\omega_E$. However, an examination of the self-consistently
determined EDOS is shown in Fig. 5, for several values of $\lambda$. Here
the collapse of the band is shown explicitly to occur for even very small
values of $\lambda$, consistent with the singular behaviour in the
real part of the self energy.
\begin{figure}[tp]
\begin{center}
\begin{turn}{-90}
\epsfig{figure=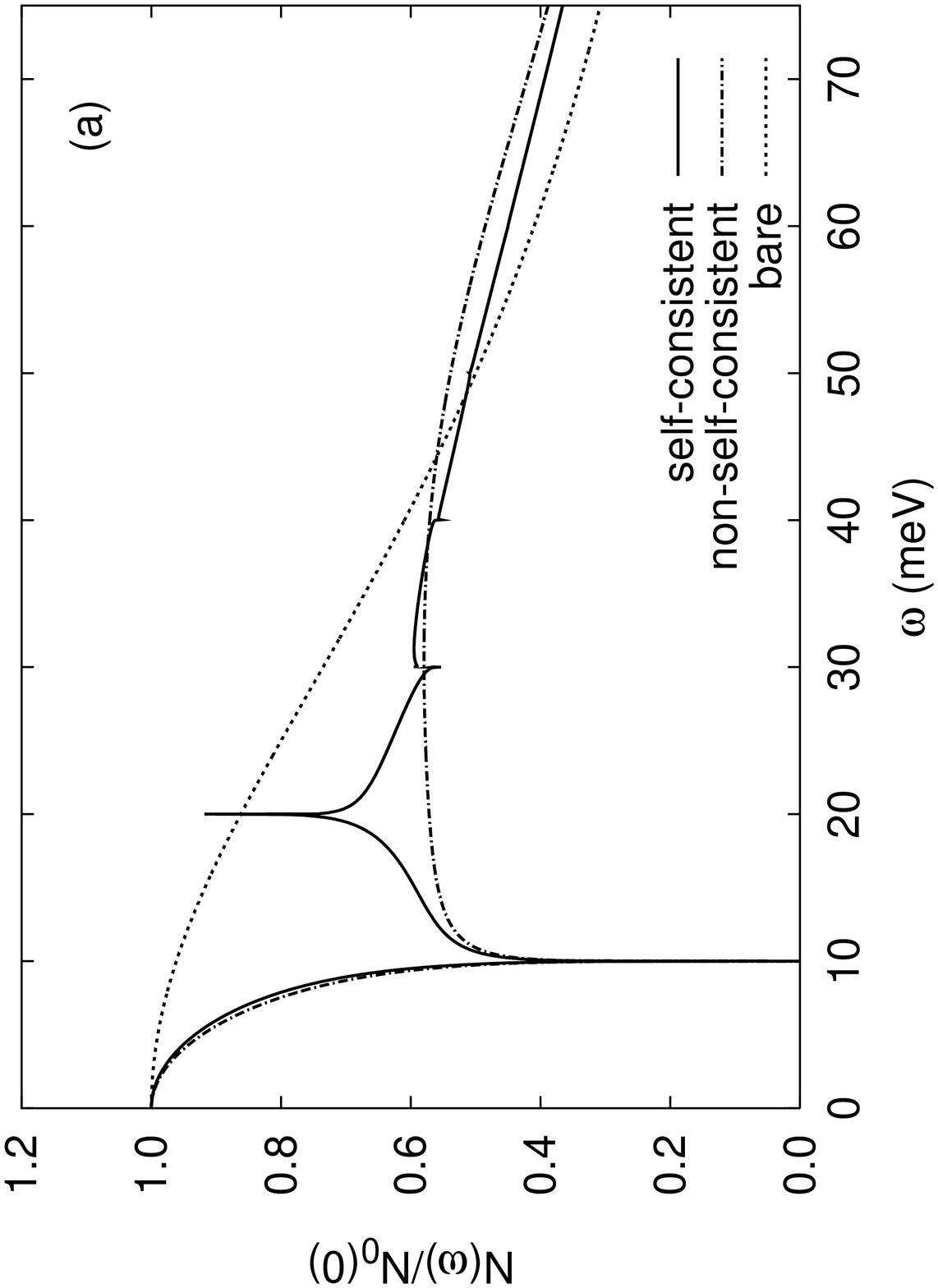,height=3.4in}
\epsfig{figure=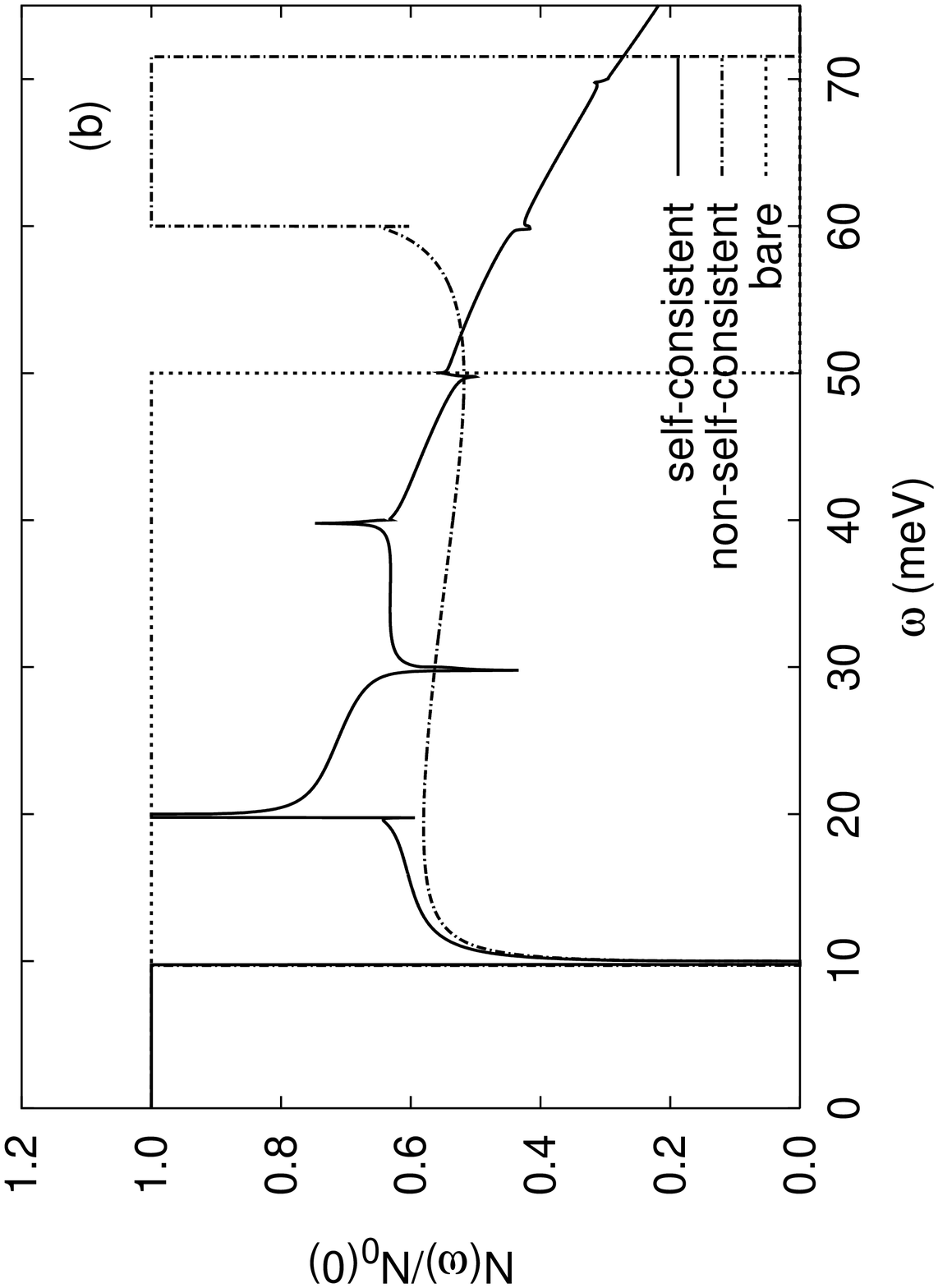,height=3.4in}
\end{turn}
\caption{ The self-consistently computed EDOS for the parameters
of Fig. 3, with (a) a Lorentzian bare EDOS, and (b) a constant
bare EDOS. Note the collapse at $\omega = \omega_E$, and the
multi-phonon structure apparent in the self-consistent
calculation. }
\end{center}
\end{figure}
Yet the Migdal approximation ought to be valid at least for very
weak coupling.

Fig. 4 shows another interesting feature, which is the significant
alteration of high energy states. This is particularly prominent in
Fig. 4b, where states are created at energies above the band edge.
This would occur for much smaller values of the electron-phonon coupling
strength, and for even larger values of the bare bandwidth, $D$. This
result is somewhat counterintuitive. We normally anticipate
that a perturbative interaction affects states just near the Fermi level.
However, here, as in the exact solution, all states are modified in an
additive way, so even states well away from the Fermi level get pushed
to higher energies.

Rather than focusing on the self energy correction itself,
Alexandrov et al. \cite{alexandrov87} used a different criterion for
the phenomenon of band collapse (which they attributed to polaron formation).
They simply took the full width at half maximum of the converged EDOS,
regardless of what structure the EDOS contains at lower frequencies. For
example, in Fig. 5 there would first be an initial {\it increase} in the
effective bandwidth as $\lambda$ increases. Only for $\lambda {{ \atop >} \atop {\sim \atop }}
3$ does the effective bandwidth decrease below $D$ (in this case
100 meV). For increased coupling strengths, the effective bandwidth
decreases smoothly to about $4D/5$, but then has a number of erratic
jumps as it decreases to approximately $2 \omega_E$. Note that Alexandrov
et al. \cite{alexandrov87} found a smooth decrease to $1\omega_E$ for two
reasons. First, much of the fine structure visible in Fig. 5 was not
obtained in their solutions, and second, they included a secondary band
which served to smooth out some of the fine structure and reduce the impact
of the electron-phonon coupling.
\begin{figure}[tp]
\begin{center}
\begin{turn}{-90}
\epsfig{figure=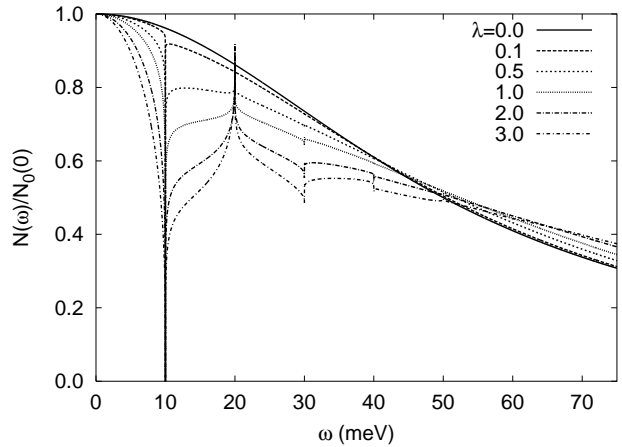,height=3.4in}
\end{turn}
\caption{ The self-consistently computed EDOS computed with a bare
Lorentzian EDOS with $\omega_E/D = 1/10$, computed for a variety
of values of the electron-phonon coupling strength. Note that the
collapse at $\omega = \omega_E$ occurs for {\it all} values of
$\lambda$. The multi-phonon structure becomes more apparent as
$\lambda$ increases. Moreover, more spectral weight is pushed to
higher energies as $\lambda$ increases. }
\end{center}
\end{figure}

Nonetheless Fig. 5 demonstrates that the Einstein model for the phonon
spectrum leads to anomalous behaviour in the self-consistent EDOS; to
determine how much of this is due to the physics of band narrowing, and
how much is attributable to the singular nature of the phonon spectrum,
we will study the effect of a broadened phonon spectrum.

\begin{figure}[tp]
\begin{center}
\begin{turn}{-90}
\epsfig{figure=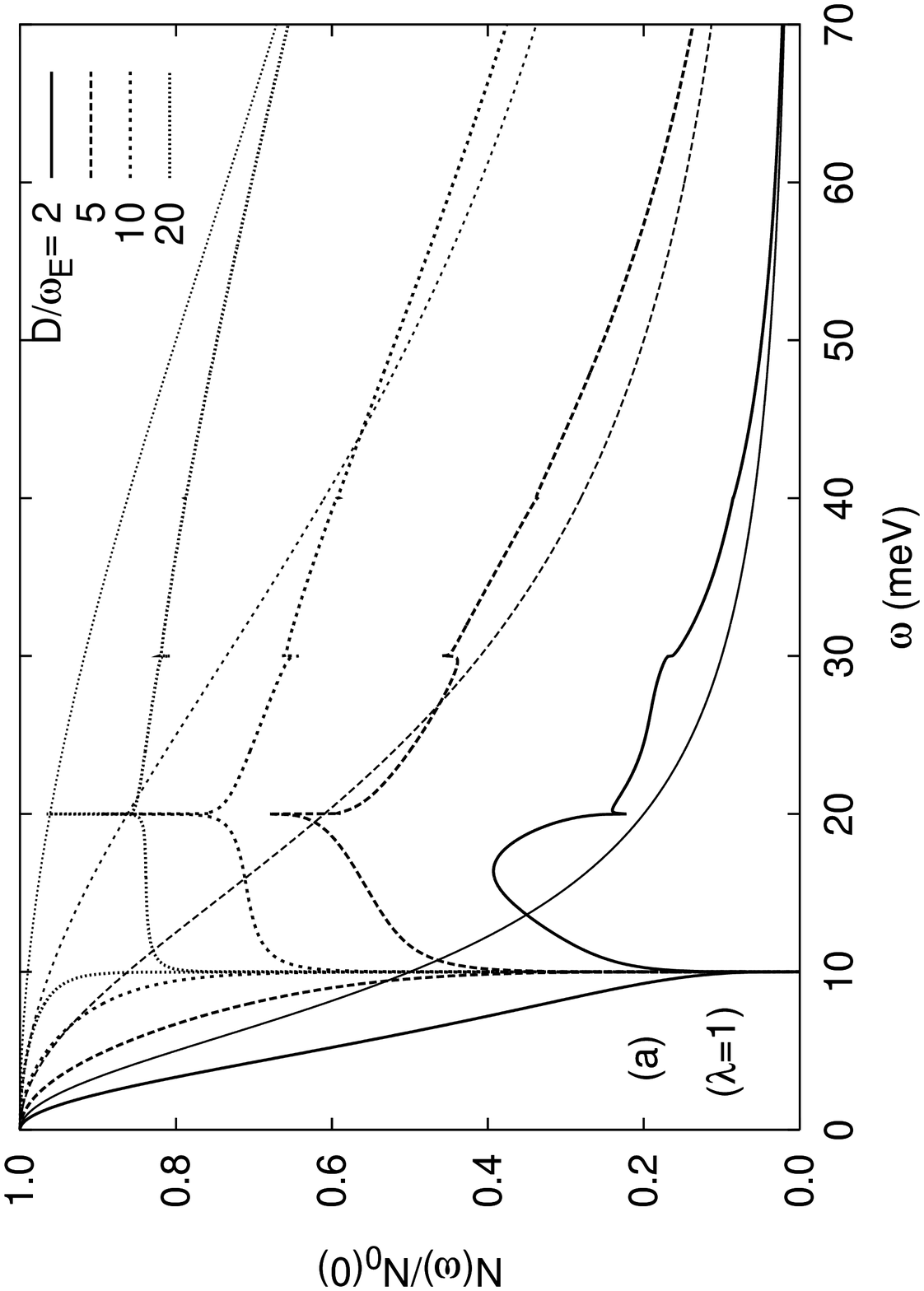,height=3.4in}
\epsfig{figure=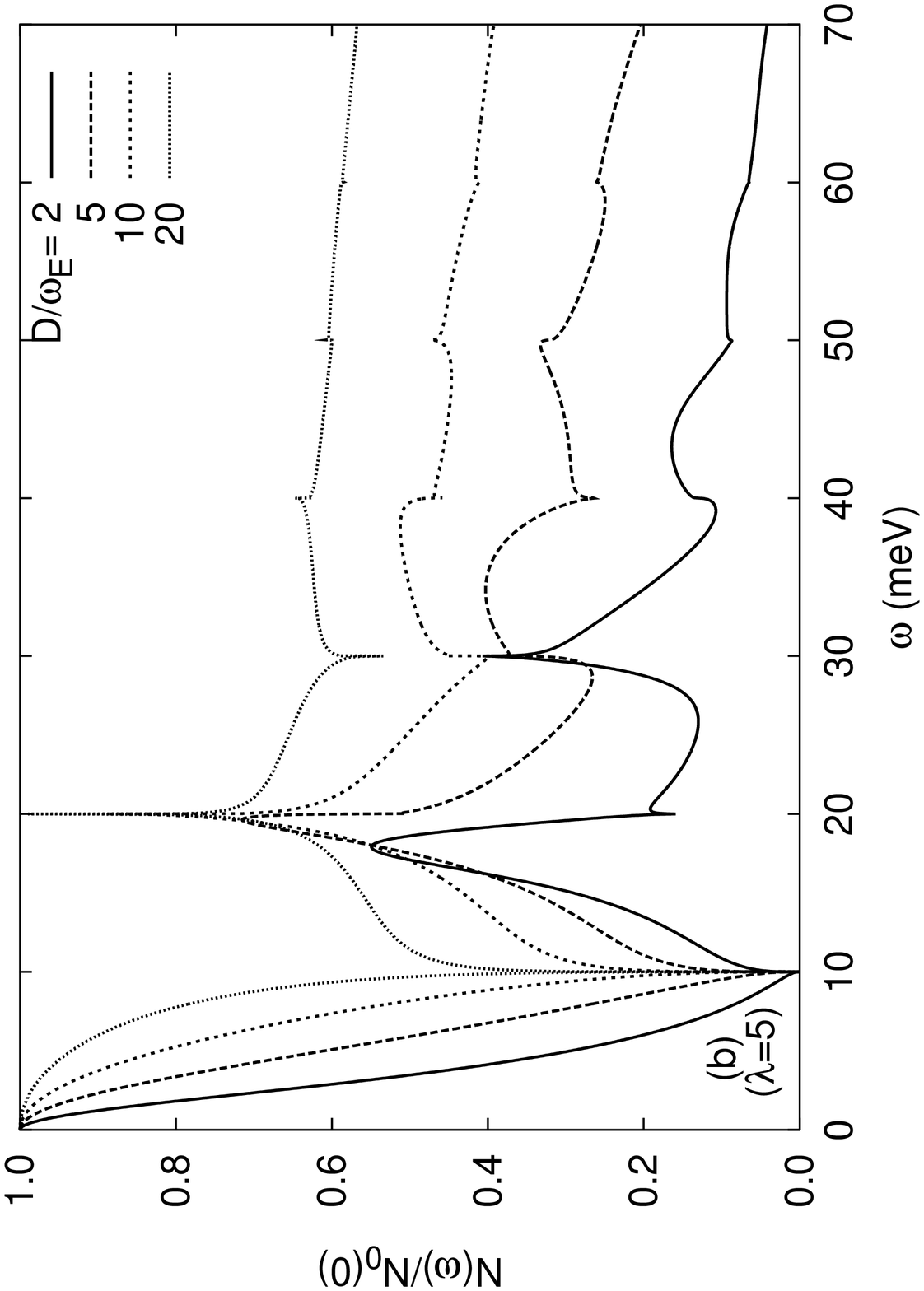,height=3.4in}
\end{turn}
\caption{ The self-consistently computed EDOS computed with a bare
Lorentzian EDOS with (a) $\lambda = 1$ and (b) $\lambda = 5$. The
latter is chosen purposely very large to better see the trends. We
use four values of the adiabatic ratio $\omega_E/D$, as labelled.
In (a) the bare Lorentzian EDOS are also shown with lighter curves
so that the modifications due to the electron-phonon interaction
are readily visible. The bare EDOS are not shown in (b) for
clarity; nonetheless, it is clear that more significant changes
occur for the higher value of $\lambda$. Note that there is not a
sum rule when the bandwidth $D$ is changed, as is the case here. }
\end{center}
\end{figure}
Before doing so, however, we show in Fig. 6 the self-consistently
calculated EDOS for several values of $\omega_E/D$. In Fig. 6a (6b) we use
$\lambda = 1.0$ (5.0), and plot the resulting EDOS for $\omega_E/D = 1/2,
1/5, 1/10$, and $1/20$. Clearly as the Einstein frequency becomes
comparable to the bare bandwidth the narrowing effects become more
pronounced, particularly for large values of $\lambda$. Interestingly,
while in the opposite limit, $\omega_E/D \rightarrow 0$, we approach
the `standard' model where the bare EDOS is unmodified by interactions,
this particular approach to that limit always shows the collapse
in the EDOS at $\omega = \omega_E$. This is true even in the case of a
constant bare band with finite width, where there is an even closer
connection to the standard model in this limit.

\noindent {\it Lorentzian phonon spectrum}

In Fig. 7 we show (a) the real part and (b) the imaginary part of
the self energy and (c) the self-consistent density of states, for
a band of electrons with a bare Lorentzian EDOS with width
$D = 10 \omega_E$ interacting with a broadened electron-phonon spectrum
($\delta = 5.0$ meV in the phonon spectrum Lorentzian centred at
$\omega_E = 10$ meV). Results are shown for $\lambda = 0,0.1,0.5,1.0,2.0$
and $3.0$. By the criterion described above that was used by
Alexandrov et al. \cite{alexandrov87}, no band narrowing has occurred
up to $\lambda = 3.0$. However, inspection of the self-consistent EDOS
shown illustrates that this criterion may be too simplistic to describe the
more global behaviour that is evident in Fig. 7.
\begin{figure}[tp]
\begin{center}
\begin{turn}{-90}
\epsfig{figure=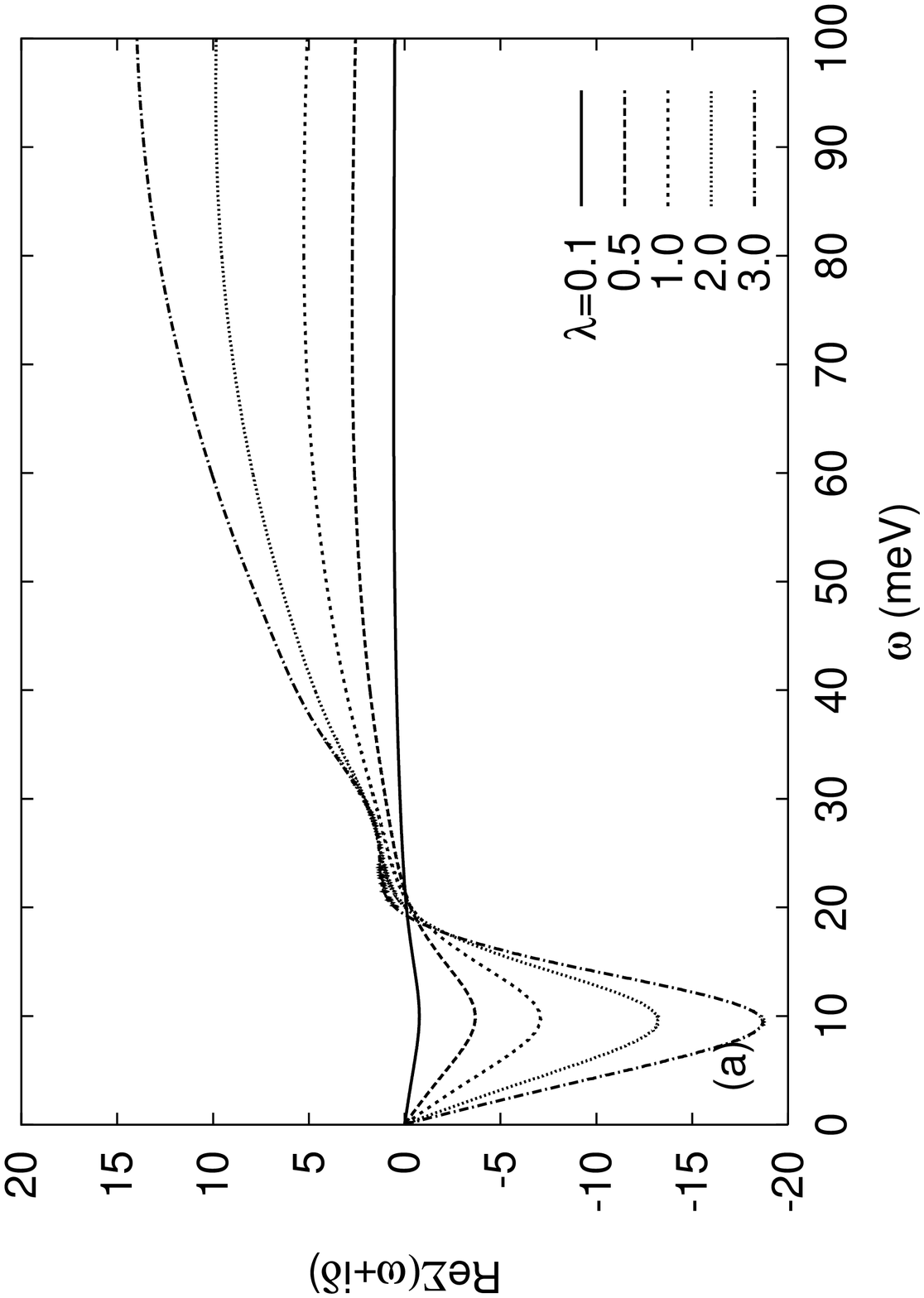,height=3.4in}
\epsfig{figure=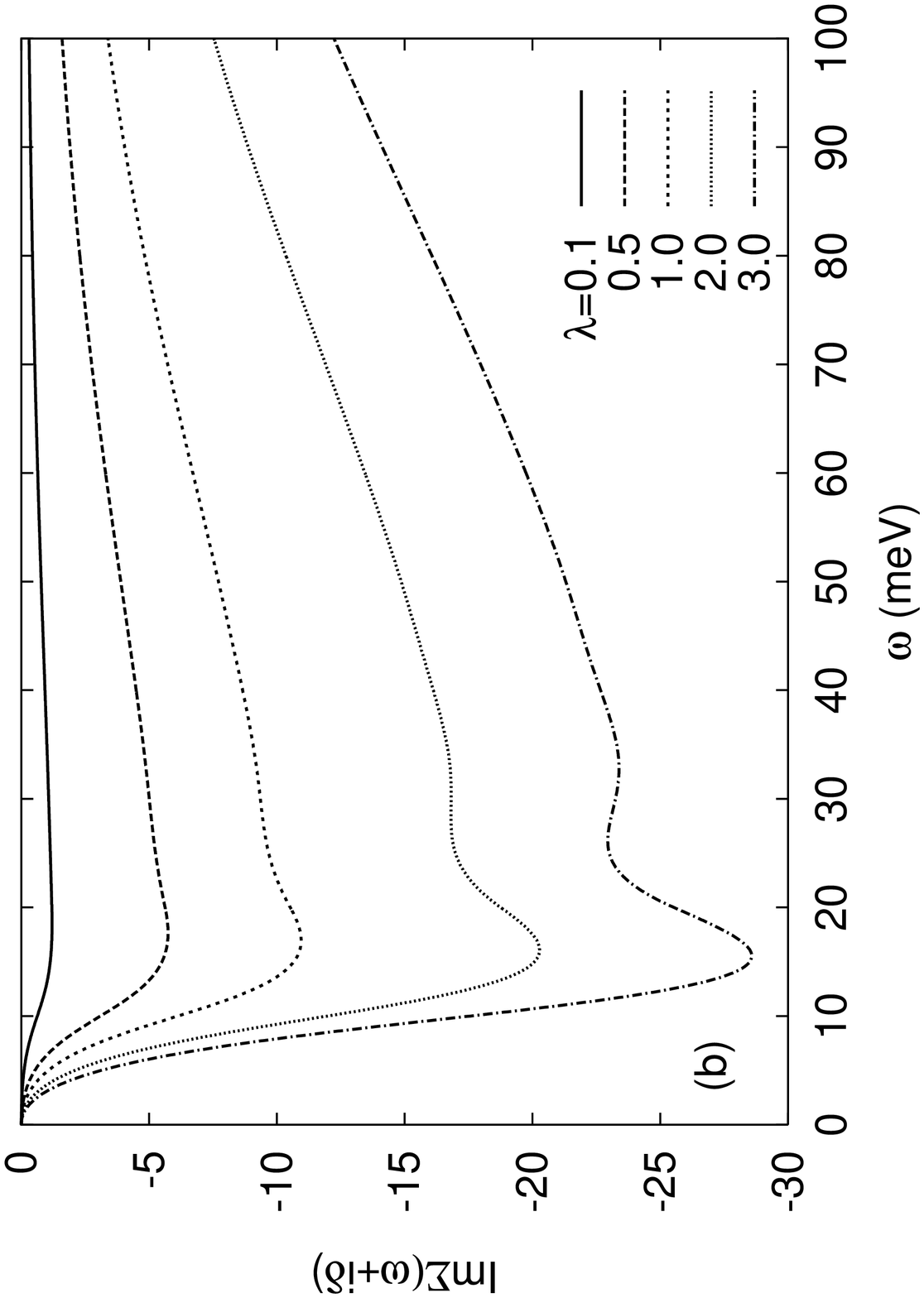,height=3.4in}
\epsfig{figure=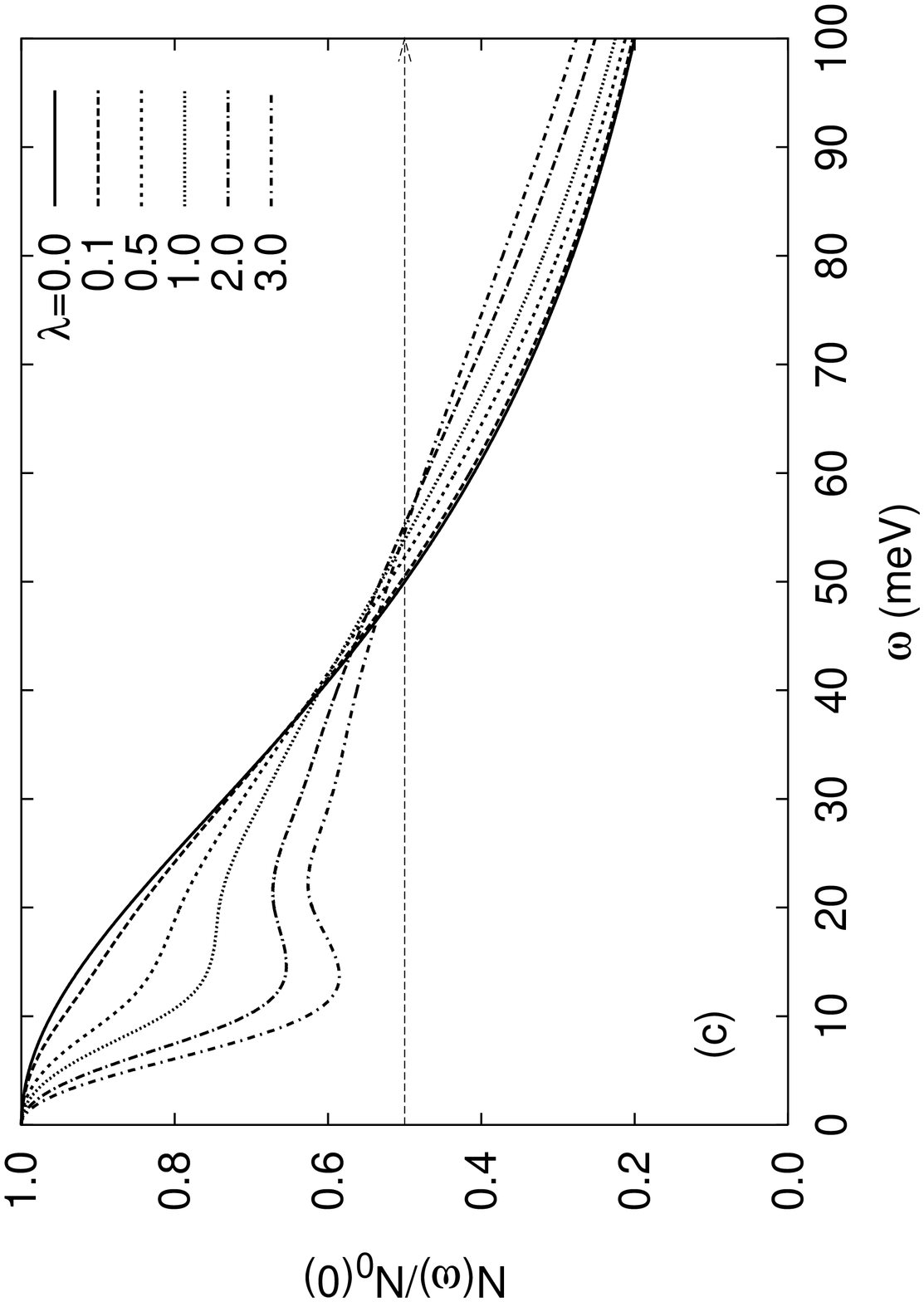,height=3.4in}
\end{turn}
\caption{ (a) Real and (b) Imaginary parts of the electron self
energy vs. frequency for a bare Lorentzian EDOS with $\omega_E/D =
1/10$, using a broadened truncated Lorentzian lineshape for the
phonon spectrum. We use central frequency $\omega_E = 10 $ meV,
halfwidth $\delta = 5$ meV, with truncation at $\eta = \omega_E
\pm 9.95$ meV. Results are shown for various values of $\lambda$
as labelled. Note that a singularity in (a) is no longer present.
Panel (c) shows the self-consistently computed EDOS using the
results of (a) and (b). Consistent with (a) a collapse at $\omega
= \omega_E$ is no longer present. All sharp features are no longer
present, but a dip near $\omega = \omega_E$ and excess spectral
weight at high energy remain. }
\end{center}
\end{figure}

For instance, it is clear that profound changes take place within $\omega_E$
of the Fermi level. So, while the full width at half maximum actually
increases with increasing $\lambda$ in the range shown, the EDOS clearly
narrows close the Fermi energy ($\omega = 0$). Thus we could plot instead
the normalized EDOS at $\omega \approx \omega_E$, for example, as a
function of coupling strength. This is shown in Fig. 8 for various values of
$\omega_E/D$.
\begin{figure}[tp]
\begin{center}
\begin{turn}{-90}
\epsfig{figure=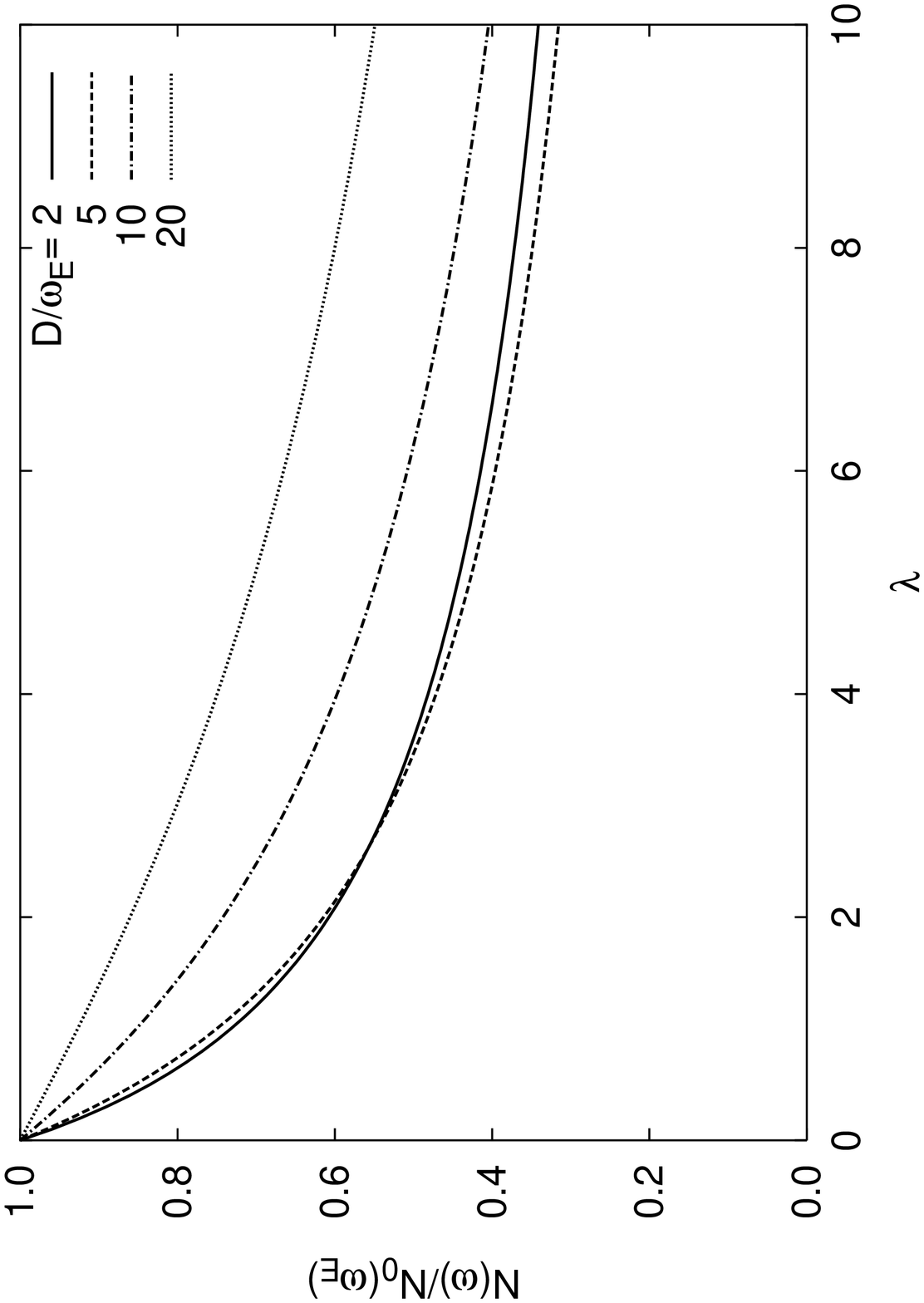,height=3.4in}
\end{turn}
\caption{ Normalized EDOS at $\omega = \omega_E$ vs. coupling
strength. This frequency is used because it roughly corresponds to
the minimum in the EDOS (see Fig. 7c). Note that the collapse is
most severe when the bare bandwidth is comparable to the
characteristic phonon frequency. Also note that the "collapse", as
measured by this indicator, evolves smoothly as a function of
coupling strength. The most significant change occurs at weak
coupling. }
\end{center}
\end{figure}

Interestingly, Fig. 8 shows that the fastest modification with increasing
coupling strength occurs near $\lambda = 0$. This is in contrast to the
criterion used in Ref. \cite{alexandrov87}, where the measure of band
collapse used there plummeted at a particular value of $\lambda \approx
3$. The development of a resonant peak near the Fermi level can indicate
new physics (e.g. polaron formation), but there is nothing in Fig. 8
to indicate that this occurs abruptly at some coupling strength.
Calculations with a bare electron band which is constant with bandwidth $D$
produce results qualitatively and quantitatively similar to those shown
in Fig. 8. This result clearly shows a very smooth evolution of the EDOS
as a function of coupling strength out to very large values of $\lambda$.
Inspection of the self-consistent EDOS as a function of frequency for
values of $\lambda$ near 10 show no qualitative differences with those
shown in Fig. 7c.

\section{SUMMARY}
\label{sec:sum}

We have revisited the Migdal approximation at a slightly more sophisticated
level than the `standard' treatment where an infinitely broad electron
band is assumed. This was done by self-consistently computing the
electron density of states for electrons in a band with finite bandwidth
interacting with phonons. In fact the self-consistency is not at all
necessary to observe the changes near the Fermi level that result from
this interaction. The non-self-consistent calculation captures the
tendency for the band to form a resonance with width given by the
characteristic phonon energy just as well. The key element is that
the bare EDOS has a finite bandwidth.

Previous work \cite{alexandrov87} has focused on the Einstein spectrum
for the phonons coupled to an electron band described by a Lorentzian
density of states. We have considered a constant bare EDOS as well, and
found very little difference in the results. A qualitative change in the
results does occur, however, when a broad phonon spectrum is used instead
of the delta function that characterizes the Einstein model. In the latter
case the electron self energy is always singular, regardless of the level
of self-consistency used to calculate the Migdal approximation. With
a bare EDOS with finite bandwidth, this singularity results in an
electron density of states that collapses at the Einstein frequency
$\omega_E$, {\it for any nonzero value of the electron-phonon coupling
constant, $\lambda$}. This collapse, however, is due to the unphysical
nature of the Einstein spectrum \cite{note5}, and does not signal a
metal-insulator transition.

As expected, the use of a broadened phonon spectrum eliminates the
singularity in the self energy, and in the self-consistent EDOS. Fig. 7c
epitomizes the change that occurs (compare with Fig. 5). There is still
a strong suppression of the EDOS at energies $\omega_E$ away from the
Fermi energy, which indicates that a resonance occurs near the Fermi level.
These states clearly are pushed to much higher energies. The fact that the
energy scale for the resonance is $\omega_E$ is indicative of increased
involvement of phonons in the electron states near the Fermi level (and
therefore of polaron formation), but there is no signal in these results
of a collapse of the conduction band for $\lambda$ of the order of unity.
Eventually, as Fig. 8 indicates, the resonance dominates the EDOS for
very large values of $\lambda$, particularly for relatively large values
of the adiabatic ratio, $\omega_E/D$. In fact, for small values of the
adiabatic ratio, the self-consistent Migdal approximation is properly
adjusting the EDOS to at least partially incorporate some of the physics
of polaron formation, i.e. that the energy scale for the electrons becomes
comparable to that of the phonons.

Ref. \cite{alexandrov87} utilized a secondary band; as the presence of this
band serves to `soften' the impact of the electron-phonon interaction on
the primary band, we have omitted it here. The same qualitative results
are obtained, except for somewhat higher values of the coupling strength
\cite{dogan02}.

Finally, over the last four decades there have been many studies of
electrons interacting with phonons, and the potential of a crossover
to a regime where polaronic behaviour dominates the physics. Many of
these studies use the Holstein model for the phonons, for the sake
of simplicity. {\bff The exact studies (see Ref. \cite{alexandrov01}
for a short review and pertinent references) account for the phonon
renormalization; thus, in principle, these include phonon broadening
effects. In practice, unfortunately, many of these studies are
carried out on finite lattices, or in various limits
(e.g. the adiabatic approximation), so that a completely satisfactory
solution is not available. Nonetheless, as reviewed in Ref. \cite{alexandrov01},
the majority of these studies suggest that a cross-over occurs from free
electron-like to polaronic behaviour, near $\lambda \approx 1$.
We make a cautionary note, however, in relation to this work; the previous
statement applies to the bare dimensionless coupling constant, $\lambda$.
However, in those studies, the `operational' value of $\lambda$ is in fact much
higher, because phonons have softened, etc. \cite{marsiglio90}. It is this
`operational' value which more closely corresponds to the value of $\lambda$
used in this work. In any event, the limitations on coupling strength in
the Migdal approximation in the normal state, or the corresponding
Eliashberg formalism in the superconducting state \cite{allen75}
must ultimately come from these exact studies. Achieving a
non-singular result within the Migdal
approximation scheme is insufficient grounds for its accuracy.}

{\bff We should also remark that some work has also been done on the
Bari\v si\'c-Labb\'e-Friedel model \cite{barisic70} (also known
as the SSH (Su-Schrieffer-Heeger) model \cite{su79}), where dispersion
in the phonon spectrum exists at the start.} This model may be distinct
from other cases where phonons are broadened due to the
electron-phonon interaction itself or anharmonicity, since it contains
sharp dispersive modes vs. individually broadened dispersive modes.
Questions concerning these models deserve further study.

\begin{acknowledgments}
This work was supported by the Natural Sciences
and Engineering Research Council (NSERC) of Canada and the Canadian
Institute for Advanced Research (CIAR).
\end{acknowledgments}

\bibliographystyle{prl}

\begin{thebibliography}{1}
\bibitem{note1} After all, the vertex correction should be compared with
a diagram {\it that is of the same order in the coupling} and is retained.
\bibitem{note2} We use the term 'Migdal approximation', although in the
literature this is often referred to as 'Migdal's Theorem'.
\bibitem{note3} {\bff Migdal \cite{migdal58} began with a bare coupling strength
$\lambda_0$, which led to strong renormalizations of both the phonon and
electron properties. In following Ref. \cite{alexandrov87} we compute electron
properties only, so that our coupling strength is already renormalized in some
sense, to the value that Migdal calls $\lambda$. While $\lambda_0$ was found to
be required to be less than unity to preserve phonon properties, $\lambda$, can,
in principle, take on larger values.}
\end{thebibliography}

\vfil\eject

\end{document}